\documentclass[review]{elsarticle}

\usepackage{lineno,hyperref}
\usepackage{amsmath}
\def\cenns{CE$\nu$NS~}
\def\cennsp{CE$\nu$NS}
\modulolinenumbers[5]

\journal{Astroparticle Physics}

\begin{document}

\begin{frontmatter}

\title{Sensitivities for coherent elastic scattering of solar and supernova neutrinos with future NaI(Tl) dark matter search detectors of COSINE-200/1T}

\author[ibs]{Young Ju Ko}
\ead{yjko@ibs.re.kr}
\author[ibs,ust]{Hyun Su Lee}
\ead{hyunsulee@ibs.re.kr}

\address[ibs]{Center for Underground Physics, Institute for Basic Science (IBS), Daejeon 34126, Republic of Korea}
\address[ust]{IBS School, University of Science and Technology (UST), Deajeon 34113, Republic of Korea}

\begin{abstract}
We investigate the prospects for measuring the coherent elastic neutrino--nucleus scattering of solar and supernova neutrinos in future NaI(Tl) dark matter detection experiments. Considering the reduced background and improved light yield of the recently developed NaI(Tl) crystals, more than 3$\sigma$ observation sensitivities of the supernova neutrino within the Milky Way are demonstrated. In the case of the solar neutrino, approximately 3$\sigma$ observations are marginal with a 1\,ton NaI(Tl) experiment assuming an order of magnitude reduced background, five photoelectron thresholds, and 5-year data exposure.
\end{abstract}


\end{frontmatter}

\section{Introduction}
A claim of dark matter observations of the DAMA/LIBRA experiment~\cite{Bernabei:2014jba,Bernabei:2018yyw} from an annual modulation of the rate of low-energy events in an array of NaI(Tl) crystals has triggered independent efforts worldwide to reproduce the annual modulation signals with these NaI(Tl) crystals~\cite{Kim:2014toa,sabre,Adhikari:2017esn,Fushimi:2018qzk,Coarasa:2018qzs,Amare:2018sxx,10.1093/ptep/ptab020,cosinus2020,PhysRevD.95.032006}. Ongoing NaI(Tl) experiments, such as COSINE-100 and ANAIS-112, have achieved background levels of approximately 2--4\,counts/kg/day/keV at the 1--6\,keV region of interest (ROI). These experiments are run with approximately 100\,kg, although DAMA/LIBRA used a 250\,kg NaI(Tl) array with a background level of less than 1\,counts/kg/day/keV in the ROI. Further efforts to reduce the background level, aiming for compatibility with DAMA/LIBRA, have been realized using high-light yields~\cite{Olivan:2017akd,Choi:2020qcj} and low-background NaI(Tl) detectors~\cite{Suerfu:2019snq,Park:2020fsq,Fushimi:2021mez}. Based on these efforts, the COSINE-200 experiment~\cite{COSINE-100:2021poy} will begin at the end of 2023 with a 200-kg NaI(Tl) detector with reduced background levels, to realize full-size and low-background NaI(Tl) crystals.


With a large number of NaI(Tl) detectors, these detectors can also be used to detect solar and supernova neutrinos via coherent elastic neutrino--nucleus scattering~(\cennsp)~\cite{Lang:2015zhv,XENON:2020gfr,Schwemberger:2021fjl}. Ton-scale dark matter detectors will soon encounter the background caused by \cenns of solar neutrino~\cite{PhysRevD.89.023524}, the so-called neutrino floor. This will be a hurdle to the dark matter searches, however, it will be also an opportunity to study new physics in the neutrino sector such as non-standard interactions~\cite{DUTTA2017242} and neutrino magnetic moment~\cite{Miranda2019}. In this study, we investigate the feasibility of future NaI(Tl) dark matter detectors as neutrino telescopes using the \cenns process.

\section{Future NaI experiments}
\label{sec_NaIExp}
\subsection{200 kg NaI(Tl) experiment (COSINE-200)}
Efforts to upgrade the ongoing COSINE-100 experiment to the next-phase COSINE-200 have resulted in the production of NaI(Tl) crystals with reduced internal backgrounds from $^{40}$K and $^{210}$Pb~\cite{Shin:2018ioq,Shin:2020bdq,COSINE:2020egt}, as well as an increased light yield of 22 photoelectrons (NPE) per unit kiloelectron volts electron-equivalent energy (keVee)~\cite{Choi:2020qcj,NEON:2022hbk}. The recrystallization method has achieved chemical purification of raw NaI powder with sufficient reduction of K and Pb contamination~\cite{Shin:2018ioq,Shin:2020bdq}. A dedicated Kyropoulos grower for small test crystals produces low-background NaI(Tl) crystals with reduced $^{40}$K and $^{210}$Pb values of less than 20\,ppb and 0.5\,mBq/kg, respectively, corresponding to background levels of less than 1\,counts/kg/day/keVee at the 1--6\,keVee ROI~\cite{COSINE:2020egt}. A full-size Kyropoulos grower was built for the 100\,kg-size crystal ingot to provide approximately 200\,kg of low-background NaI(Tl) detectors for the COSINE-200 experiment. The expected background levels of these crystals were less than 0.5\,counts/kg/day/keVee in the ROI~\cite{COSINE-100:2021poy}. This estimation was based on the measured background levels of the small test crystals reported in Ref.~\cite{COSINE:2020egt}.

An increased light yield of the NaI(Tl) crystal is essential to reduce the energy threshold below a keVee. With an optimized concentration of thallium doping in the crystal, we achieved a high light yield of 17.1$\pm$0.5\,NPE/keVee, which is slightly larger than that of the COSINE-100 crystal (approximately 15\,NPE/keVee). A further increase in the light-collection efficiency by $\sim$50\% in the NaI(Tl) crystal was achieved by an improved encapsulation scheme, as described in Ref.~\cite{Choi:2020qcj}. In this scheme, the crystal and photomultiplier tube (PMT) are directly connected without an intermediate quartz window, for which a 22\,NPE/keVee light yield is achieved. A similar crystal encapsulation technique was applied to the reactor \cenns search experiment NEON, and approximately 22\,NPE/keVee light yields were measured~\cite{NEON:2022hbk}.

The typical trigger requirement of the COSINE-100 experiment is satisfied by coincident photoelectrons in two PMTs attached to each side of the crystal at approximately 0.13\,keVee. However, PMT-induced noise events are dominantly triggered at energies below a few keVees. The multivariable boosted decision tree (BDT) provided a 1\,keVee analysis threshold with less than 0.1\% noise contamination and above 80\% selection efficiency~\cite{Adhikari:2020xxj}. A key variable in the BDT is the likelihood parameter using the event shapes of scintillation-like and PMT-induced noise-like events. Further improvement of the low-energy event selection is ongoing using the COSINE-100 data by developing new parameters for the BDT and employing a machine learning technique that uses raw waveforms directly. COSINE-200 targets an analysis threshold of 5\,NPE~(0.2\,keVee)~\cite{COSINE-100:2021poy}, which is similar to the energy threshold that has already been achieved by the COHERENT experiment with CsI(Na) crystals~\cite{COHERENT:2017ipa} and the target threshold of the NEON \cenns search experiment with NaI(Tl) crystals~\cite{NEON:2022hbk}.

\begin{figure}[!htb]
\begin{center}
\includegraphics[width=0.95\columnwidth]{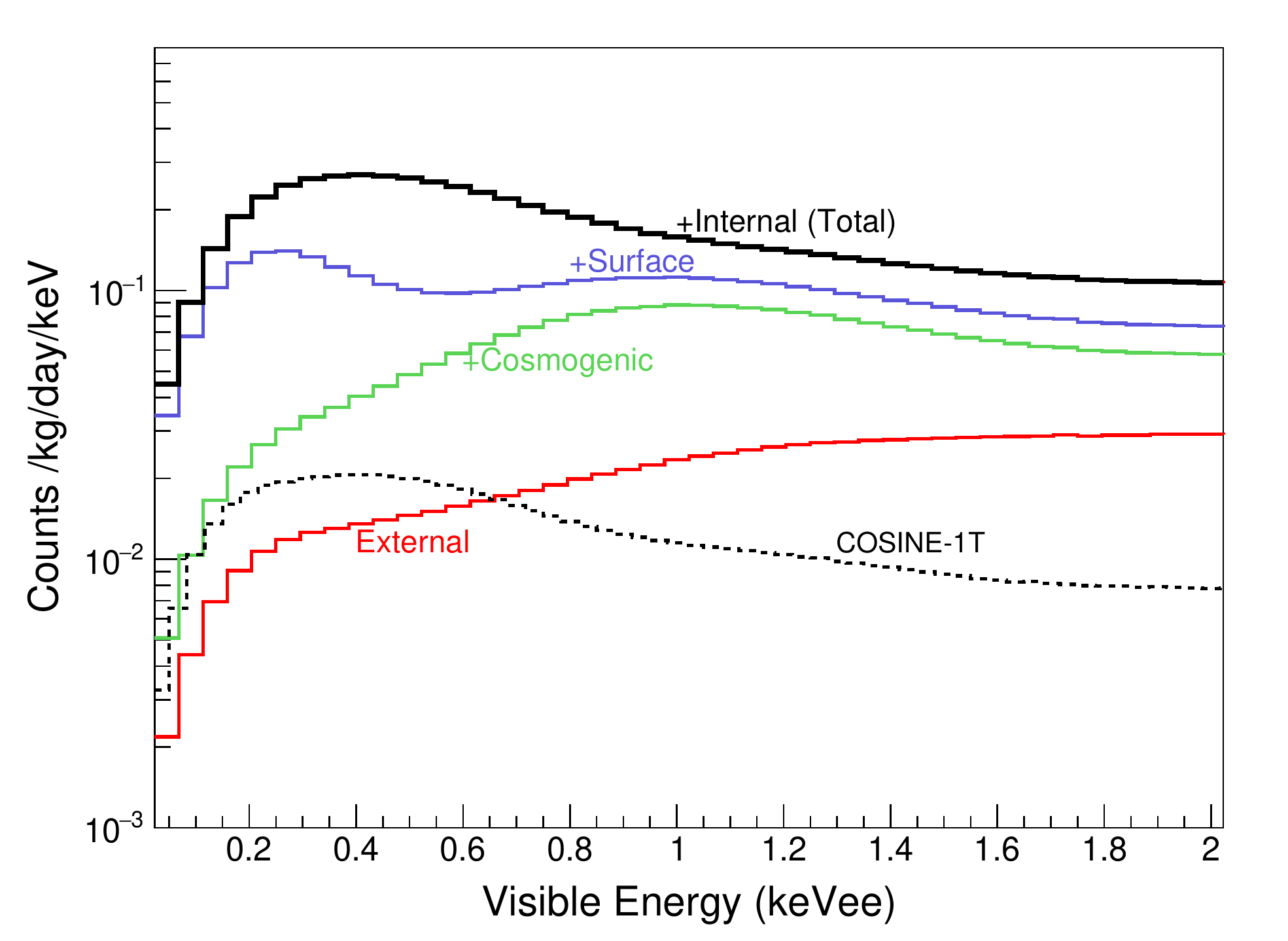} 
\caption{The black solid line is the expected background spectra of the COSINE-200 experiment while the black dotted line is the target background of the COSINE-1T experiment. The expected spectrum of COSINE-200 is based on the developed low-background NaI(Tl) crystal~\cite{COSINE:2020egt}. The red line denote external backgrounds due to the radiations from outside the crystal such as PMT and acrylic table. The green line is the summation of external and cosmogenic components by $^3$H, $^{113}$Sn and $^{109}$Cd, {\it{etc}}. The blue line shows spectrum adding the surface background to the green line, which is caused by $^{210}$Pb contaminated on the crystal surface. The solid black line representing the total background is the addition of the internal background to the blue line, which is caused by radioactivity inside the crystal contaminated with U/Th/K and $^{210}$Pb.}
\label{fig_exp}
\end{center}
\end{figure}

The COSINE-200 experiment can be realized in a 4$\times$4 array of 12.5\,kg NaI(Tl) modules by replacing the crystals inside the COSINE-100 shield~\cite{Adhikari:2017esn}. The COSINE-200 experiment will run for at least 3 years for an unambiguous test of DAMA/LIBRA annual modulation signals~\cite{Adhikari:2015rba}. In addition to the verification of the DAMA/LIBRA experiment, this experiment can achieve the best sensitivity for low-mass dark matter searches, especially for spin-dependent WIMP--proton interactions~\cite{COSINE-100:2021poy}. Sensitivities on \cenns from the solar and supernova neutrinos have background-level assumptions, as shown in Fig.~\ref{fig_exp}, 22\,NPE/keVee high-light yield~\cite{NEON:2022hbk}, and 5-year data.

\subsection{1000 kg NaI experiment (COSINE-1T)}
Assuming the successful operation of the 200\,kg NaI detectors and achieving the best sensitivities for low-mass dark matter of the WIMP-proton spin-dependent interaction, it is normal to consider a large-scale NaI(Tl) dark matter search experiment corresponding to COSINE-1T (1000\,kg NaI). Multiple developments in high-quality NaI(Tl) detectors have been extensively investigated to reduce the background and increase the light yield. The responses of the NaI(Tl) crystal at $-$35$^{\circ}$C showed an increased light yield of approximately 5\% in electron-equivalent energy and an additional 10\% increase in $\alpha$-induced events~\cite{Lee:2021aoi}. Silicon photomultipliers (SiPMs) can replace conventional PMTs owing to their increased quantum efficiency and reduced radioactive background~\cite{Baudis_2018,Lee:2021jfx}. If we use SiPMs, pure NaI without thallium doping can be optional and operated at liquid nitrogen temperature, and this option is expected to provide improved high light yield~\cite{PureNaI1,absoluteLY,Ding:2022jjm}.

In the estimation of the COSINE-1T sensitivity for the \cennsp, we consider an improved light yield of 30\,NPE/keVee and reduced background. Assuming an improvement in the internal background reduction on raw powder purification and crystallization, as well as removing external backgrounds dominated by attached PMTs, our target background is  an order of magnitude reduction compared to the COSINE-200 crystal in the ROI. However, we also evaluate the sensitivities assuming  the same background of the COSINE-200 considering ambiguity of the background reduction for the COSINE-1T experiment.  The threshold and exposure are used for sensitivity estimation with the same values as for the 200-kg NaI experiment (2--8\,NPE threshold and 5-year data).

\section{Neutrino sources}
\label{sec_nusrc}
Since its prediction in 1974~\cite{PhysRevD.9.1389}, searching for the \cenns from various neutrino sources was conducted without success until the first observation in 2017 by COHERENT collaboration~\cite{COHERENT:2017ipa}. They used neutrinos from a spallation neutron source with an energy of approximately 30 MeV. Relatively high-energy neutrinos with significant background reduction using the timing information of the pulsed beam allowed the first observation of \cennsp. However, such success has not been achieved using other neutrino sources, such as reactor, solar, and supernova neutrinos.

The sun is the strongest constant source of neutrinos, regardless of the specific location or time on Earth. Solar neutrinos have energies similar to reactor neutrinos; therefore, they are difficult to observe. The observation of the \cenns from the solar neutrino can be expected to contribute to solar physics through information on the flux of solar neutrinos. In addition, low-energy \cenns measurements can provide complementary information to \cenns experiments using accelerator. Considering the solar neutrino flux and its energies, large-size and low-energy threshold detectors are required. These requirements may be satisfied by ton-scale low-background dark matter search detectors~\cite{Drukier19842295,PhysRevD.91.095023,XENON:2020gfr}.

Supernovae are the most intensive, but transient, sources of neutrinos in the universe~\cite{Janka:2017vlw}. Because of huge amount of neutrinos in a short time period of less than 10\,s, the \cenns from the supernova neutrinos occurring in the Milky way can be observed by low-background dark matter search detectors~\cite{Lang:2015zhv,DarkSide20k:2020ymr,Fushimi:2021kce}. The flux of supernova neutrinos for energy and time through observations of supernova \cenns can provide interesting information on supernova properties.



\subsection{Solar neutrino}
\begin{figure*}[!htb]
\begin{center}
\includegraphics[width=0.49\columnwidth]{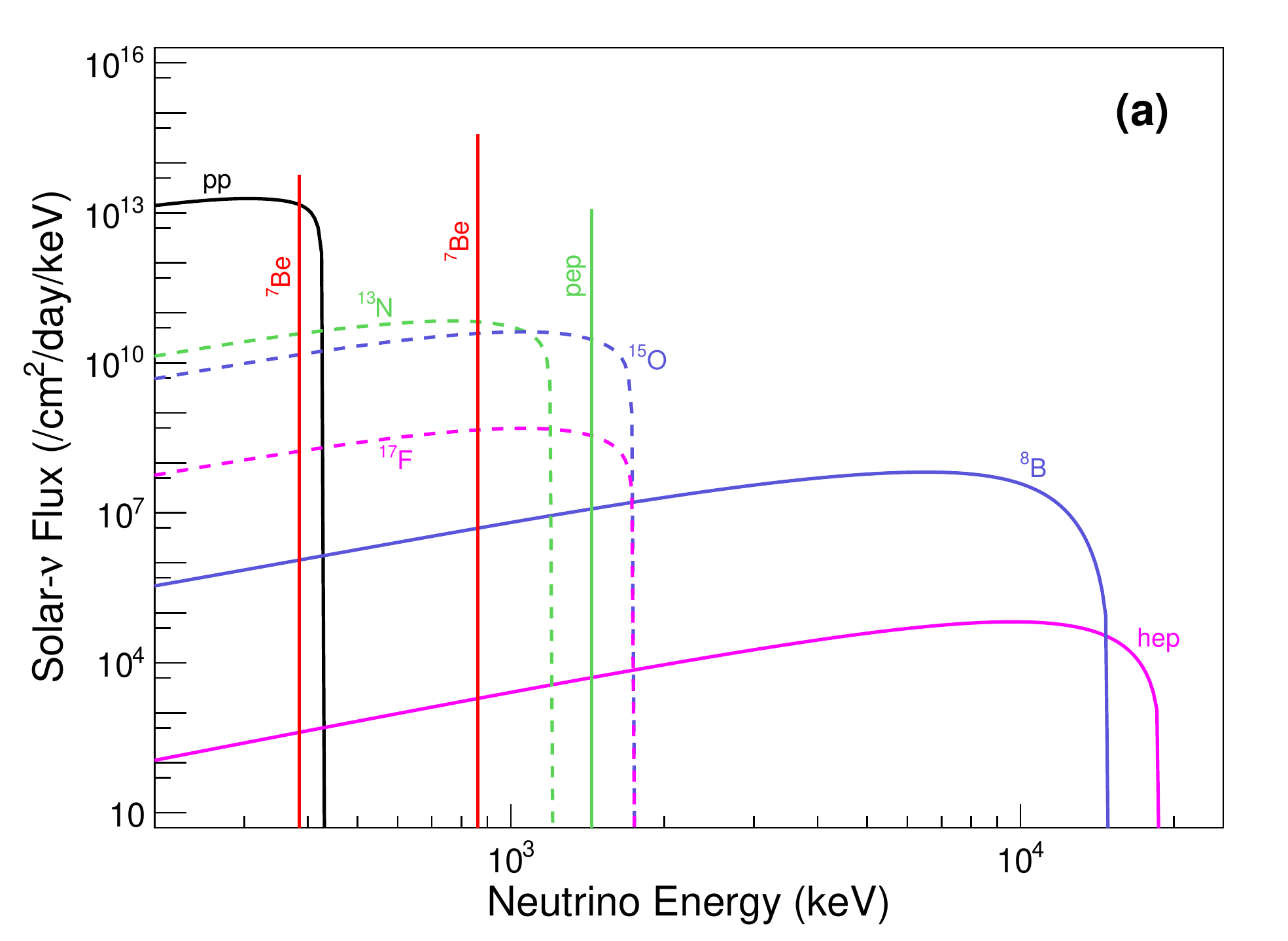} 
\includegraphics[width=0.49\columnwidth]{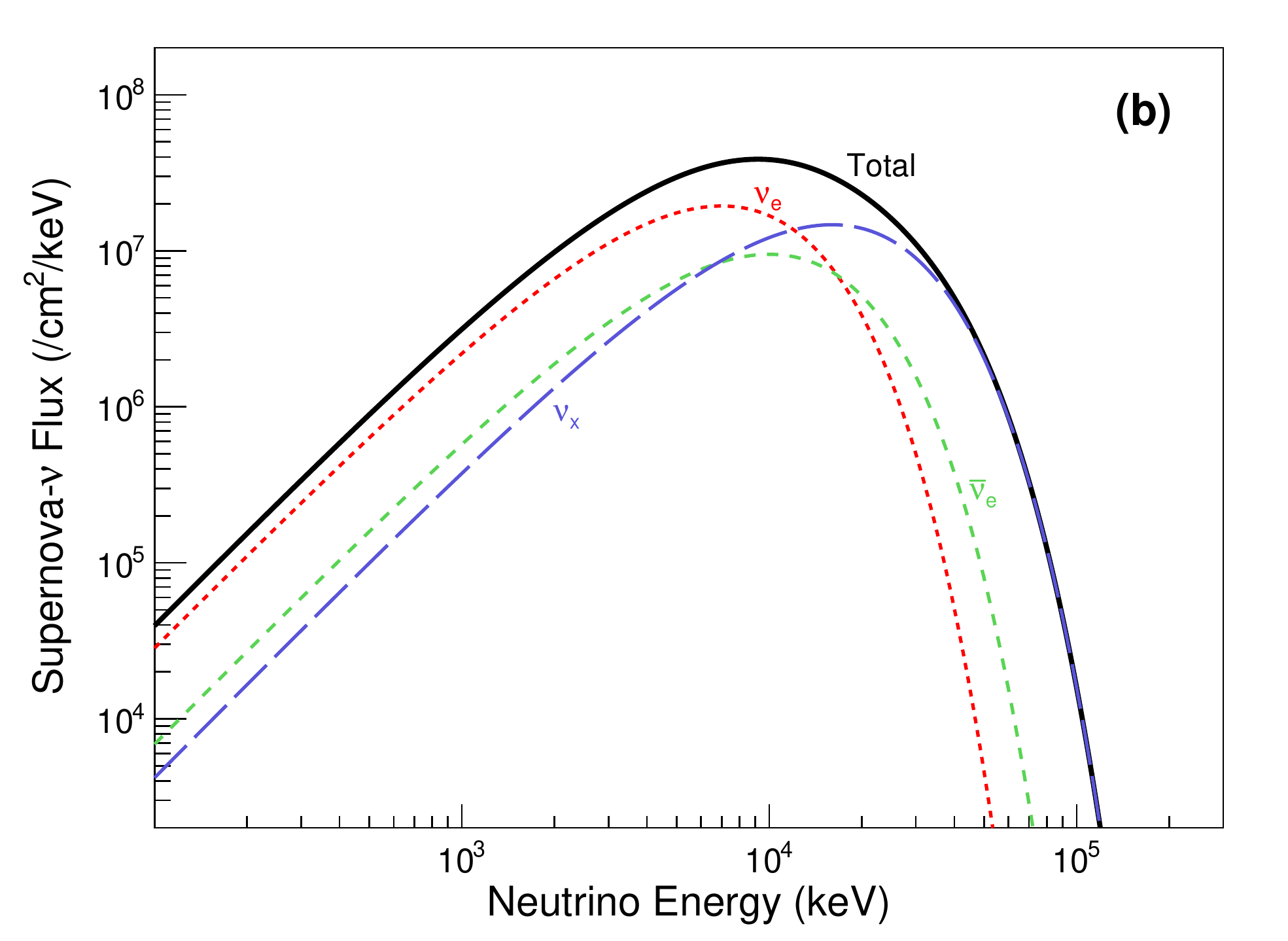} 
\caption{Energy spectra for the solar neutrino~\cite{Bahcall_2005} and the supernova neutrinos~\cite{Biassoni_2012}. (a) Solid and dashed lines indicate the neutrino sources from the proton--proton chain reaction and the CNO cycle, respectively. (b) The distance to supernova is assumed to be 10\,kpc, and $\nu_x$ (green dotted line) shows the summed flux of neutrino flavors, except for $\nu_e$ and $\bar{\nu}_e$.}
\label{fig_flux}
\end{center}
\end{figure*}

Since the first observation of solar neutrinos in the Homestake mine~\cite{solar_homestake}, several theoretical and experimental studies have improved our understanding of solar neutrinos~\cite{IANNI201444}. Fusion processes via the proton-proton chain and carbon-nitrogen-oxygen (CNO) cycle emit neutrinos. Figure~\ref{fig_flux}~(a) shows the solar neutrino spectra~\cite{Bahcall_2005} from various sources expected by a standard solar model~\cite{1989ApJ.339.1156G}. The total neutrino flux is approximately $7\times10^{10}\,$/cm$^2$/s. Among the various sources, solar $^{8}$B neutrinos contribute the dominant signals for \cenns in the NaI(Tl) detectors.

\subsection{Supernova neutrino}
A core collapse supernova, which occurs at the end of a star's life with masses heavier than eight solar masses, is an interesting source of neutrinos for \cenns observations. The neutrino flux at the detector $\Phi^d_i$ from the supernova can be approximated by Boltzmann distribution~\cite{Biassoni_2012} as follows:
\begin{eqnarray}
\Phi^d_i(E_\nu) = \frac{1}{4\pi L^2}\,\Phi^s_i(E_\nu) = \frac{1}{4\pi L^2}\,\frac{N_i}{2T_i^3}\,E_\nu^2\,\exp\left(-\frac{E_\nu}{T_i}\right),
\end{eqnarray}
where $\Phi^s_i$ is the neutrino emission spectrum, $E_\nu$ is the neutrino energy, and $L$ is the distance from the supernova to the detector (Earth). Further, $T_i$ is the temperature of the emitted neutrinos for flavor $i$, and $k_B$ is the Boltzmann constant. The values of $k_B T_i$ are 3.5, 5, and 8\,MeV for $\nu_e$, $\bar{\nu}_e$, and $\nu_x$, respectively, \cite{Biassoni_2012}.  Here, $\nu_x$ are all neutrino flavors, without $\nu_e$ and $\bar{\nu}_e$ ( $\nu_{\mu}$, $\bar{\nu}_{\mu}$, $\nu_{\tau}$ and $\bar{\nu}_{\tau}$). Further, $N_i$ is the number of emitted neutrinos. The total released energy $E_{rel}$ is related to the emission spectrum as follows:
\begin{eqnarray}
E_{rel} = 6\cdot\int_0^\infty E_\nu\cdot\Phi^s_i(E_\nu)\,dE_\nu.
\end{eqnarray}
Factor 6 was derived from a simplified model assuming the equipartition of $E_{rel}$ ($\sim3\times10^{53}$\,ergs) for the six neutrino flavors during production~\cite{Biassoni_2012}. Figure~\ref{fig_flux}~(b) shows the expected neutrino energy spectra from a supernova explosion that occurs at a distance of 10\,kpc from Earth.

\section{Coherent elastic neutrino--nucleus scattering from NaI(Tl) dark matter search experiments}
The differential rate for \cenns in terms of the recoil energy $E_r$ is given by
\begin{eqnarray}
\frac{dR}{dE_r} =\sum_t^\mathrm{Na,I} n_t\int_{E_\nu^{\mathrm{min}}} dE_\nu\,\Phi(E_\nu)\,\frac{d\sigma_t}{dE_r},
\label{eq_rate}
\end{eqnarray}
where $n_t$ is the number of target nuclei and $\Phi(E_\nu)$ is the neutrino flux described in Sec.~\ref{sec_nusrc} for each source. The minimum energy of the neutrino required for a target nucleus to recoil $E_\nu^\mathrm{min}$ is $\sqrt{E_rM_t/2}$, where $M_t$ is the mass of the target nucleus. The differential cross-section $d\sigma_t/dE_r$ can be written as follows~\cite{Freedman:1977xn}
\begin{eqnarray}
\frac{d\sigma_t}{dE_r} = \frac{G_F^2M_t}{2\pi} \left[(G_V + G_A)^2 + (G_V-G_A)^2 \left(1-\frac{E_r}{E_\nu}\right)^2 - (G_V^2-G_A^2)\frac{E_rM_t}{2E_\nu^2}  \right],
\label{eq:xcfull}
\end{eqnarray}
where $G_V$ and $G_A$ are coefficients related with vector and axial-vector coupling, respectively, and $G_F$ is the Fermi constant. 
Considering tiny contribution of axial term~\cite{Hoferichter:2020osn,PhysRevD.98.053004}, the differential cross section can be approximated as 
\begin{eqnarray}
\frac{d\sigma_t}{dE_r} = \frac{G_F^2M_t}{8\pi}\left[Z(4\sin^2\theta_\mathrm{W} - 1) + N\right]^2\left(2-\frac{E_rM_t}{E_\nu^2}\right)|f(E_r)|^2,
\label{eq_crossection}
\end{eqnarray}
where $\theta_\mathrm{W}$ is the weak mixing angle, and $Z$ and $N$ are the numbers of protons and neutrons in the target nucleus, respectively. The nuclear form factor $|f(q)|^2$ can be approximated as the Helm form factor~\cite{PhysRev.104.1466,Lewin:1995rx}.

\begin{figure*}[!htb]
\begin{center}
\includegraphics[width=0.49\columnwidth]{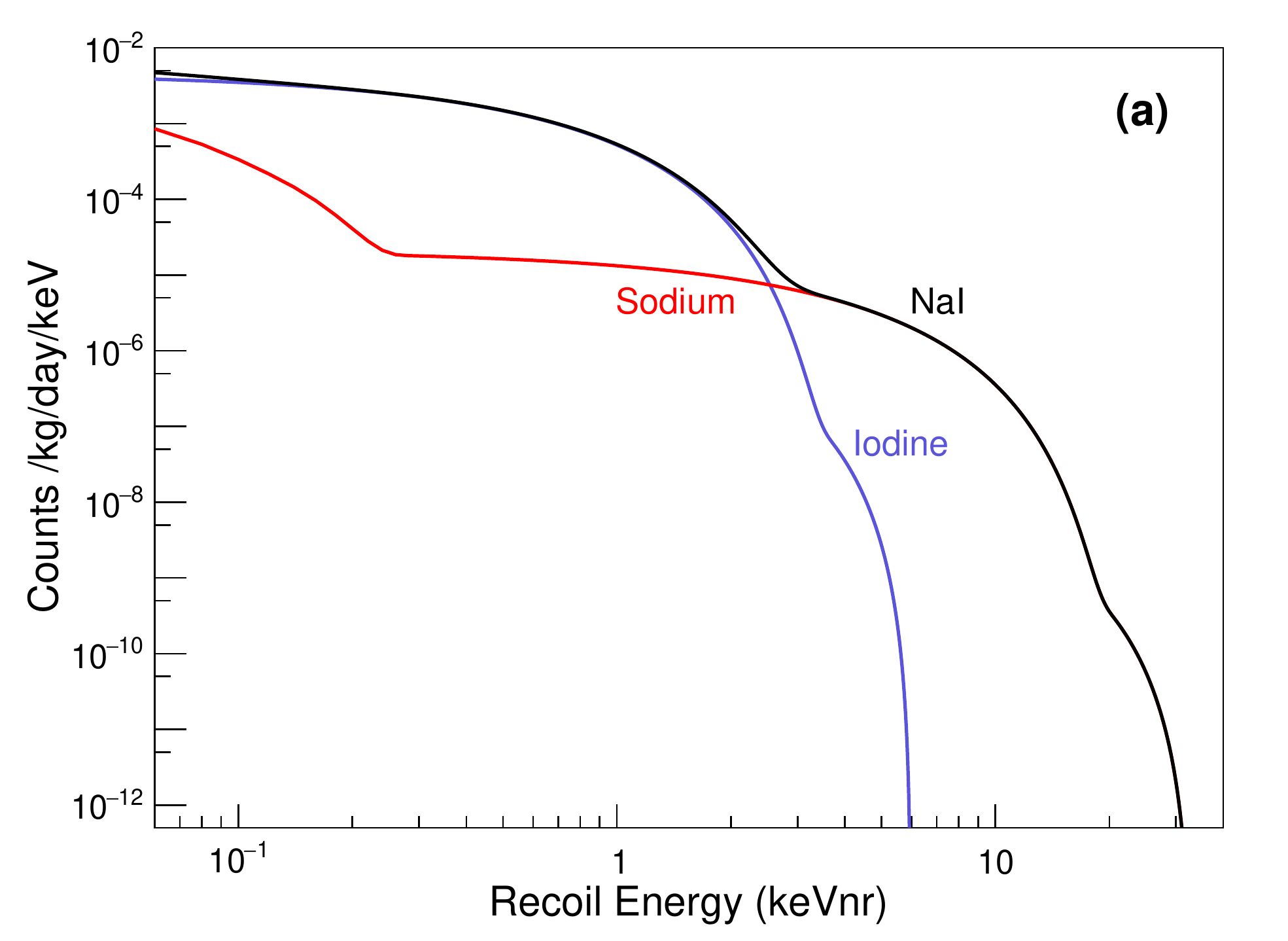}
\includegraphics[width=0.49\columnwidth]{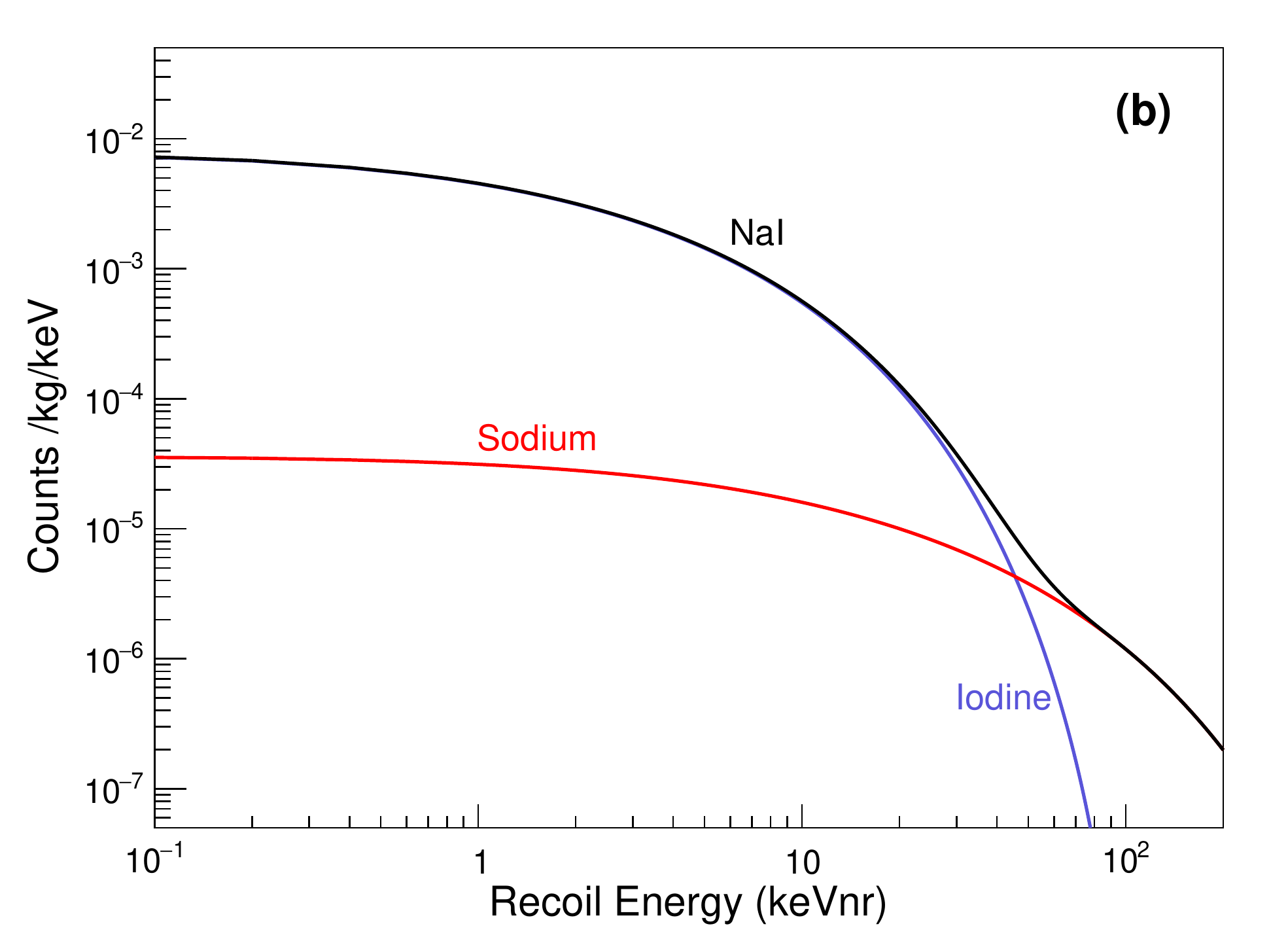}
\caption{Expected nuclear recoil energy spectra of \cenns from the (a) solar neutrino and (b) supernova neutrino interacted with NaI(Tl) crystals are presented. Sodium (red color) and iodine (blue color) recoils are separately indicated. }
\label{fig_cevns}
\end{center}
\end{figure*}

Figures~\ref{fig_cevns} (a) and (b) show the expected \cenns in kiloelectron volt nuclear recoil (keVnr) energy for the solar and supernova neutrinos, respectively. The low mass number of sodium nuclei provides a higher energy deposition, but the cross-section is enhanced by $N^2$ in Eq.~\ref{eq_crossection}. Therefore, iodine is the dominant target for  \cenns observations in both solar and supernova neutrinos.

\subsection{Detector response to generate signals}
To express the \cenns rate in terms of the electron-equivalent visible energy, the effect of detector response should be considered. The detector response comprises quenching and scintillation processes in the NaI(Tl) crystal, generating photoelectrons and amplifying electrons in the PMT, and triggering and digitizing the signal waveform. Here, we performed a fast simulation to estimate the effect of the detector response.

\begin{figure}[!htb]
\begin{center}
\includegraphics[width=0.95\columnwidth]{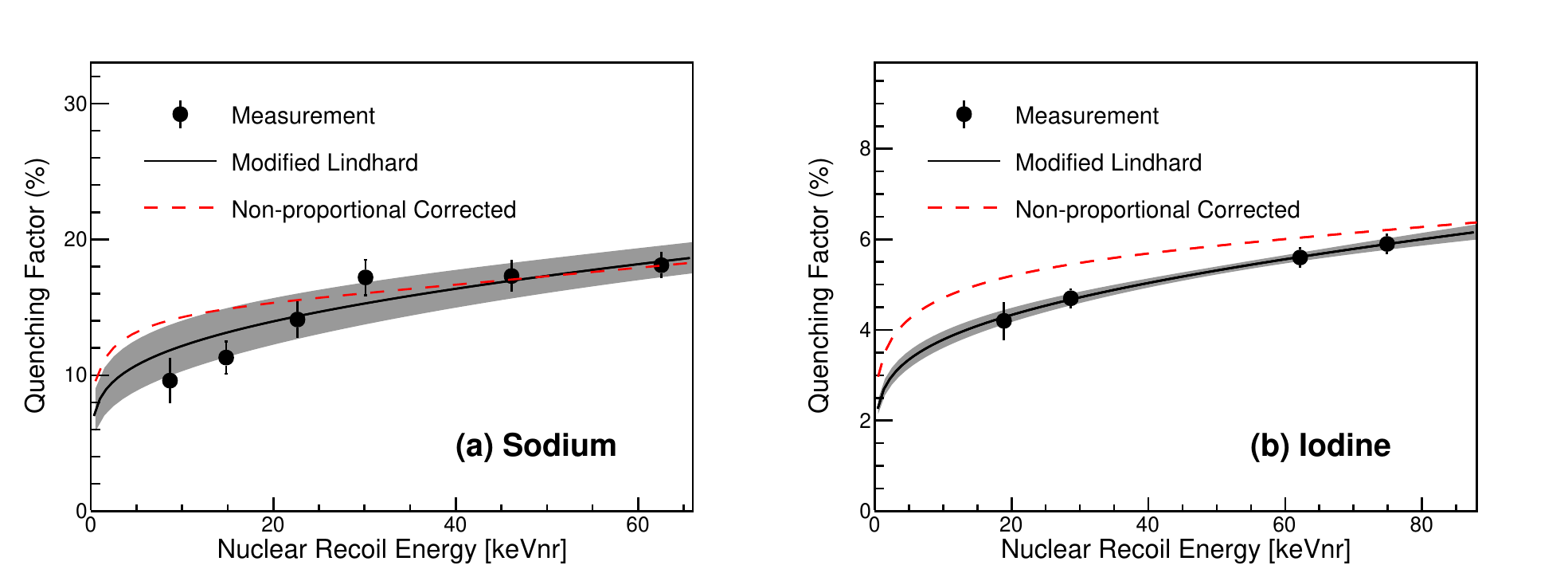} 
\caption{Quenching factors for (a) sodium and (b) iodine. Black dots are measurements~\cite{Joo:2018hom}, and black solid lines are modeled with the measurements in Ref.~\cite{Ko:2019enb}. Red dashed lines are the corrected model for non-proportionality.}
\label{fig:qf}
\end{center}
\end{figure}

For an event with nuclear recoil energy $E_r$, the number of photoelectrons from the NaI(Tl) crystal attached to the two PMTs can be modeled as follows:
\begin{eqnarray}
f(N_\mathrm{pe}|E_r,L) = f_{Pois}\left[N_\mathrm{pe}|\mu_\mathrm{pe} = Q_t(E_r)\cdot E_r\cdot L\right]
= \frac{(\mu_\mathrm{pe})^{N_\mathrm{pe}}}{N_\mathrm{pe}!}e^{-\mu_\mathrm{pe}},
\label{eq:poisson}
\end{eqnarray}
where $f_{Pois}$ is the Poisson distribution with Poisson mean $\mu_\mathrm{pe}$, $Q_t(E_r)$ is the quenching factor (QF) of the target material $t$ for the nuclear recoil energy $E_r$, and $L$ is the light yield per unit visible energy (keVee), as discussed in Sec.~\ref{sec_NaIExp}. The QF is the scintillation light yield for nuclear recoil relative to that for electron/$\gamma$-induced radiation of the same energy. Since the QF was recently measured above recoil energies of 8.7\,keV and 18.9\,keV for sodium and iodine, respectively~\cite{Joo:2018hom}, they were modeled using the modified Lindhard model~\cite{osti_4701226} shown in Fig.~\ref{fig:qf}, as described in Refs.~\cite{Ko:2019enb}, for extrapolation to lower energies. The QF measurements used a calibration method of the 59.54\,keV line from $^{241}$Am source, assuming a linear response of the NaI(Tl) crystal. However, because of the nonproportionality of the NaI(Tl) crystal~\cite{nonprop}, background modeling of the NaI(Tl) crystals considers a nonproportional calibration function, as described in Ref.~\cite{cosinebg2}, which was also applied to the background spectra in Fig.~\ref{fig_exp}. Various calibration methods lead to different QF results, as discussed in Ref.~\cite{Cintas:2021fvd}. Therefore, we applied a correction of the non-proportionality to the QF, as shown in Fig.~\ref{fig:qf}, for the following evaluations.

The number of photoelectrons $N_\mathrm{pe}$ is observed by two (left and right) PMTs attached to the two ends of the crystal, and $N_\mathrm{pe}$ for each PMT is distributed as a binomial distribution:
\begin{eqnarray}
f(N_\mathrm{pe}^\mathrm{left}|N_\mathrm{pe},~p) = f_B(N_\mathrm{pe}^\mathrm{left}|N_\mathrm{pe},~p)
= \begin{pmatrix} N_\mathrm{pe}\\N_\mathrm{pe}^\mathrm{left} \end{pmatrix} p^{N_\mathrm{pe}^\mathrm{left}} (1-p)^{N_\mathrm{pe} - N_\mathrm{pe}^\mathrm{left}},
\label{eq:binomial}
\end{eqnarray}
where $f_B$ is the binomial distribution, $N_\mathrm{pe}^\mathrm{left}$ is the number of photoelectrons in the left PMT, and $p$ is the probability of producing photoelectrons in the left PMT when a photoelectron is generated. The $p$ is equivalent to the ratio of the light yield for the left PMT to the total light yield. We assume $p = 0.5$, which is the same probability for each PMT. In the fast simulation, $N_\mathrm{pe}$ and $N_\mathrm{pe}^\mathrm{left}$ were randomly generated based on Eqs.~\ref{eq:poisson} and \ref{eq:binomial} for each event with recoil energy $E_r$ and the number of photoelectrons for the right PMT can be obtained as $N_\mathrm{pe}^\mathrm{right} = N_\mathrm{pe} - N_\mathrm{pe}^\mathrm{left}$. These are the inputs to simulate the scintillation waveform for each PMT. 

The time of every single photoelectron (SPE) generated in the simulation, $t_{spe}$, was randomly extracted from a reference waveform of the NaI(Tl) crystal, which is an accumulation of 59.54\,keV $\gamma$ events from the $^{241}$Am source~\cite{Kim:2014toa}. We assumed that the time distribution of each SPE had a Gaussian shape. The size of the SPE ($A_{spe}$) is determined by the amplification of the PMT with $N_{step}$ dynode amplification steps and total amplification factor $\tilde{N}_{Amp}$. Each stage amplification of $n_{A/s}$ electrons is modeled with a Poisson distribution with Poisson mean $\mu_{A/s}$, 
\begin{eqnarray}
\mu_{A/s} = \left(\tilde{N}_{Amp}\right)^{\frac{1}{N_{step}}}.
\end{eqnarray}
The number of electrons for $i^\mathrm{th}$ step is
\begin{eqnarray}
n_{e,i} = \sum_{j=0}^{n_{e,i-1}} n_{A/s},~~~i>0,
\end{eqnarray}
where $n_{A/s}$ is randomly generated, based on the Poisson distribution for each $j$. This process was performed independently for each SPE.

\begin{figure}[!htb]
\begin{center}
\includegraphics[width=0.95\columnwidth]{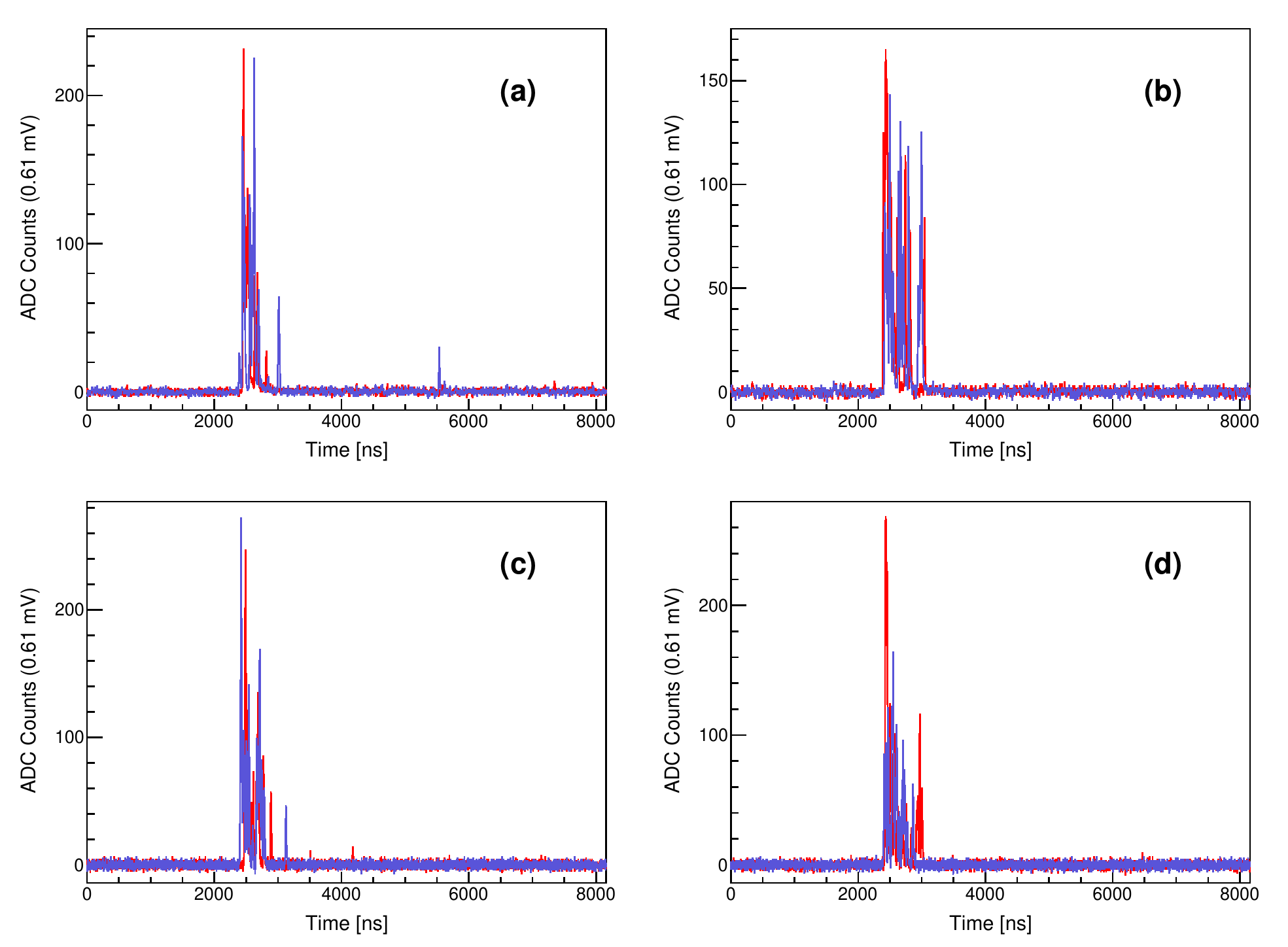} 
\caption{Examples for data waveforms (a and b) and simulation waveforms (c and d). Red and blue colors denote different PMTs (left and right, respectively). The energy range of the examples is from 1 to 1.5\,keV.}
\label{fig:waveform}
\end{center}
\end{figure}

\begin{figure*}[!htb]
\begin{center}
\includegraphics[width=0.49\columnwidth]{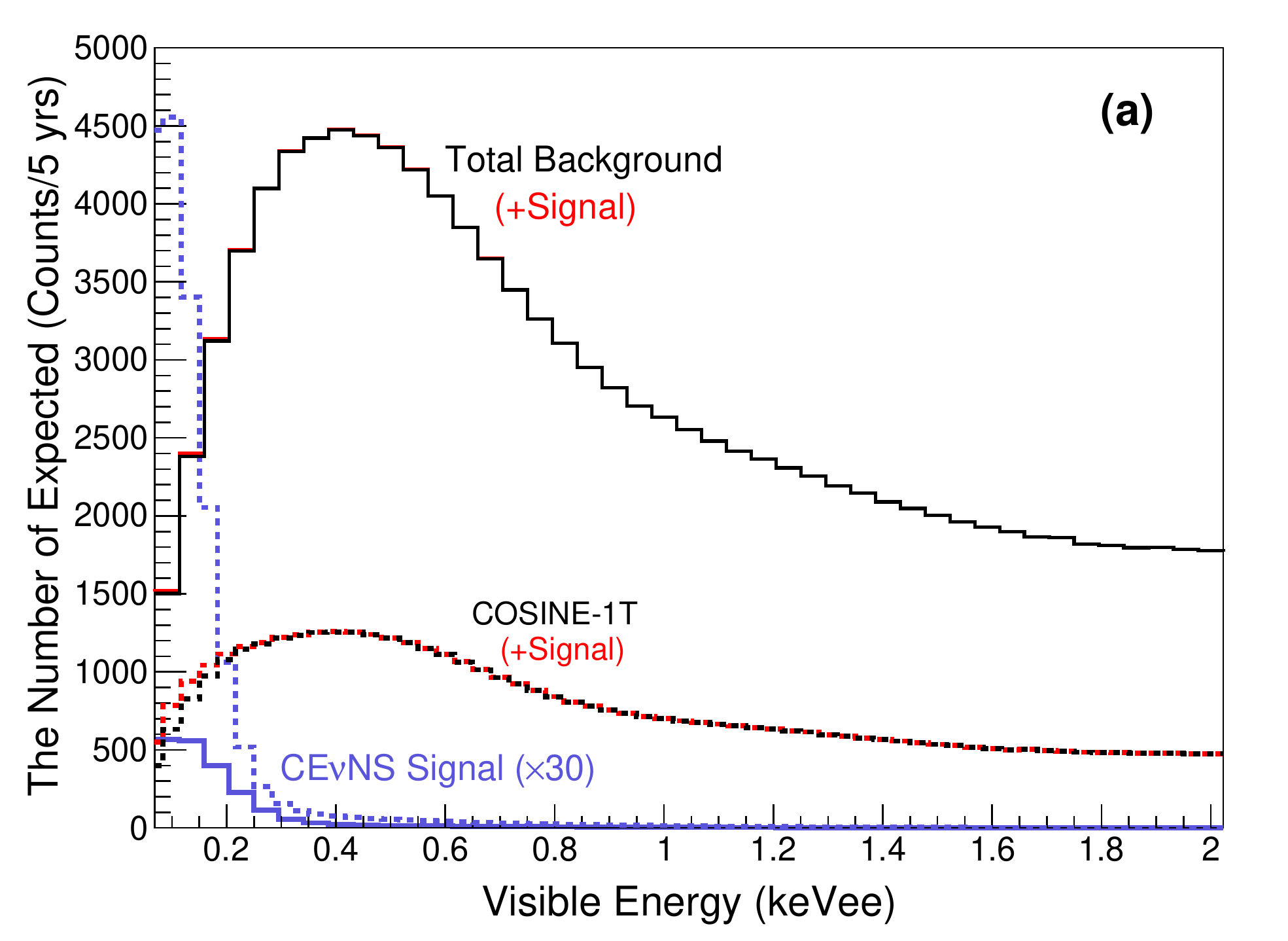} 
\includegraphics[width=0.49\columnwidth]{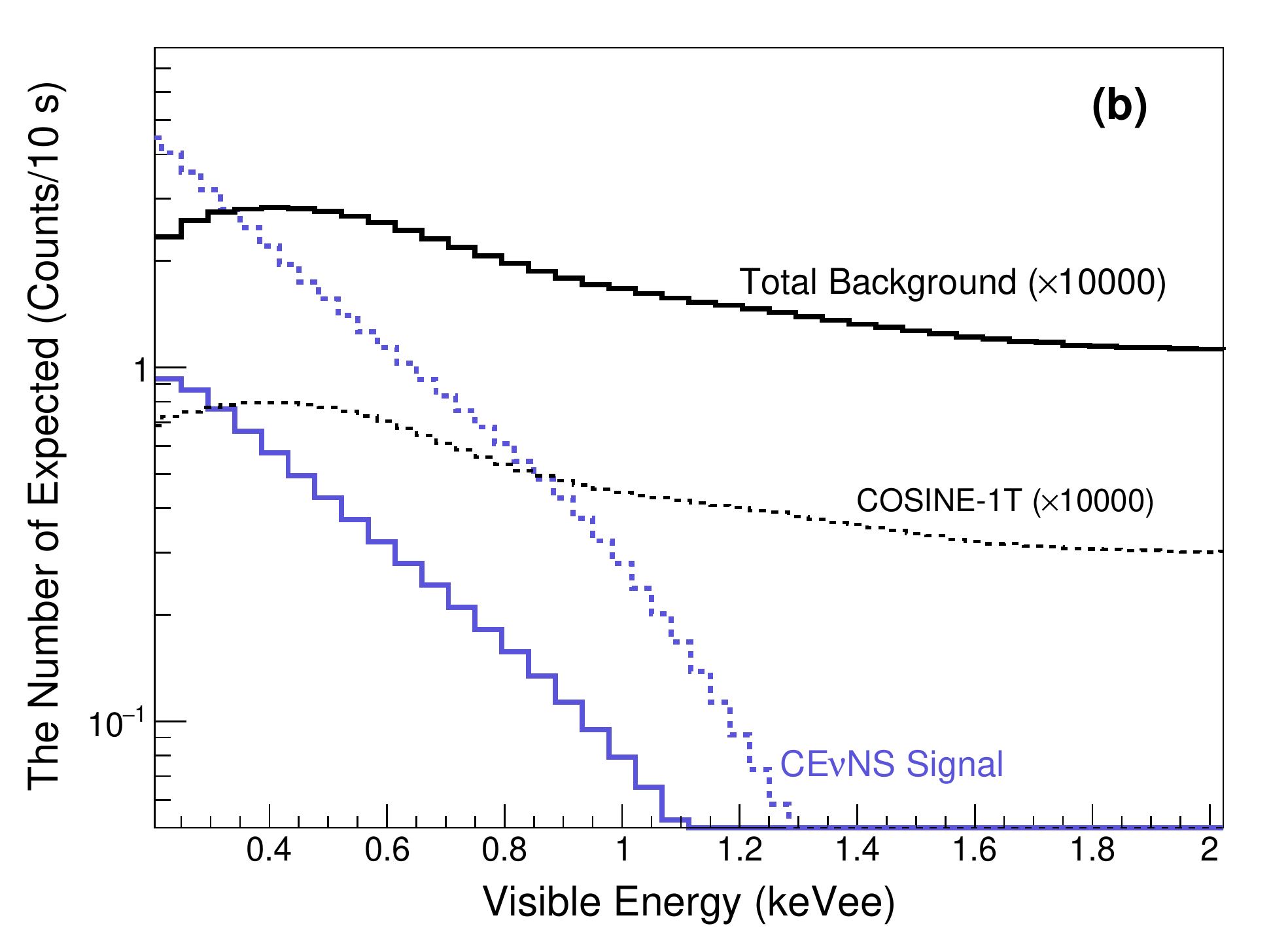} 
\caption{Visible spectra of the \cenns (blue) for the (a) solar neutrino and (b) supernova neutrino with the expected background spectra (black). The solid and dotted lines are for 200-kg and 1000-kg NaI(Tl) crystals, respectively, and the red lines show the background added by the \cenns signal. The exposure time is 5\,years and 10\,s for (a) and (b), respectively, and the distance of the supernova is assumed to be 10\,kpc.}
\label{fig_visE}
\end{center}
\end{figure*}

The waveform of each time bin ($w(t)$) can then be written as follows: 
\begin{eqnarray}
w^\mathrm{left/right}(t) = ped + \sum_{i=1}^{N_\mathrm{pe}^\mathrm{left/right}} C_{qd}\,q_e\,A_{spe}\cdot f_{Gaus}(t|t_{spe},\sigma_{spe}),
\end{eqnarray}
where $ped$ is the pedestal, $C_{qd}$ is a conversion factor for the charge to ADC counts related to the data acquisition system, $q_e$ is the charge of an electron, and $f_{Gaus}$ is the Gaussian shape of the SPE. The standard deviation $\sigma_{spe}$ is obtained from the data. The pedestal is also measured from the data as its mean and standard deviation, which provide random generation assuming a Gaussian shape. This waveform was digitized based on a specification of 500-MHz sampling rate, 12-bit resolution, and a peak-to-peak dynamic range of 2.5\,V, which was used for the COSINE-100 experiment~\cite{Adhikari:2018fpo}. The last stage was triggered via the same trigger logic used in COSINE-100 data acquisition~\cite{Adhikari:2018fpo}. Fig.~\ref{fig:waveform} presents an example of the waveforms from the data and simulation. In this process, we can produce simulated energy spectra for both the backgrounds and signals (Fig.~\ref{fig_visE}).

\subsection{Sensitivity for \cenns}
To evaluate the sensitivities of \cenns observations from different neutrino sources, binned maximum likelihood fits to the simulated energy spectra  were performed. The likelihood is built using Poisson probability distributions:
\begin{eqnarray}
L(\mu) = \prod_i \frac{(\mu s_i + b_i)^{n_i}}{n_i!}e^{-(\mu s_i + b_i)},
\label{eq:likelihood}
\end{eqnarray}
where $s_i$ denotes the number of expected \cenns events, $b_i$ denotes the number of expected background events, and $n_i$ denotes the number of observed events in the $i^\mathrm{th}$ energy bin. The number of \cenns signals when $\mu = 1$ was obtained from the visible energy-based signal spectra (Fig.~\ref{fig_visE}). The null hypothesis is evaluated by $\mu = 0$, and a statistical hypothesis test for the presence of \cenns signals is performed by comparing the signal hypothesis to the null hypothesis as follows:
\begin{eqnarray}
t_0 = -2\ln\frac{L(\mu=0)}{L(\hat{\mu})},
\label{eq:testS0}
\end{eqnarray}
where $\hat{\mu}$ denotes the fitted value of $\mu$ for the maximum likelihood fit of  $L(\mu)$. The test statistic $t_0$ is equivalent to the $\chi^2$ difference between the best-fit signal ($\mu = \hat{\mu}$) and the null signal ($\mu = 0$), assuming a Gaussian distribution of the data~\cite{Workman:2022ynf}.

\begin{figure*}[!htb]
\begin{center}
\includegraphics[width=0.49\columnwidth]{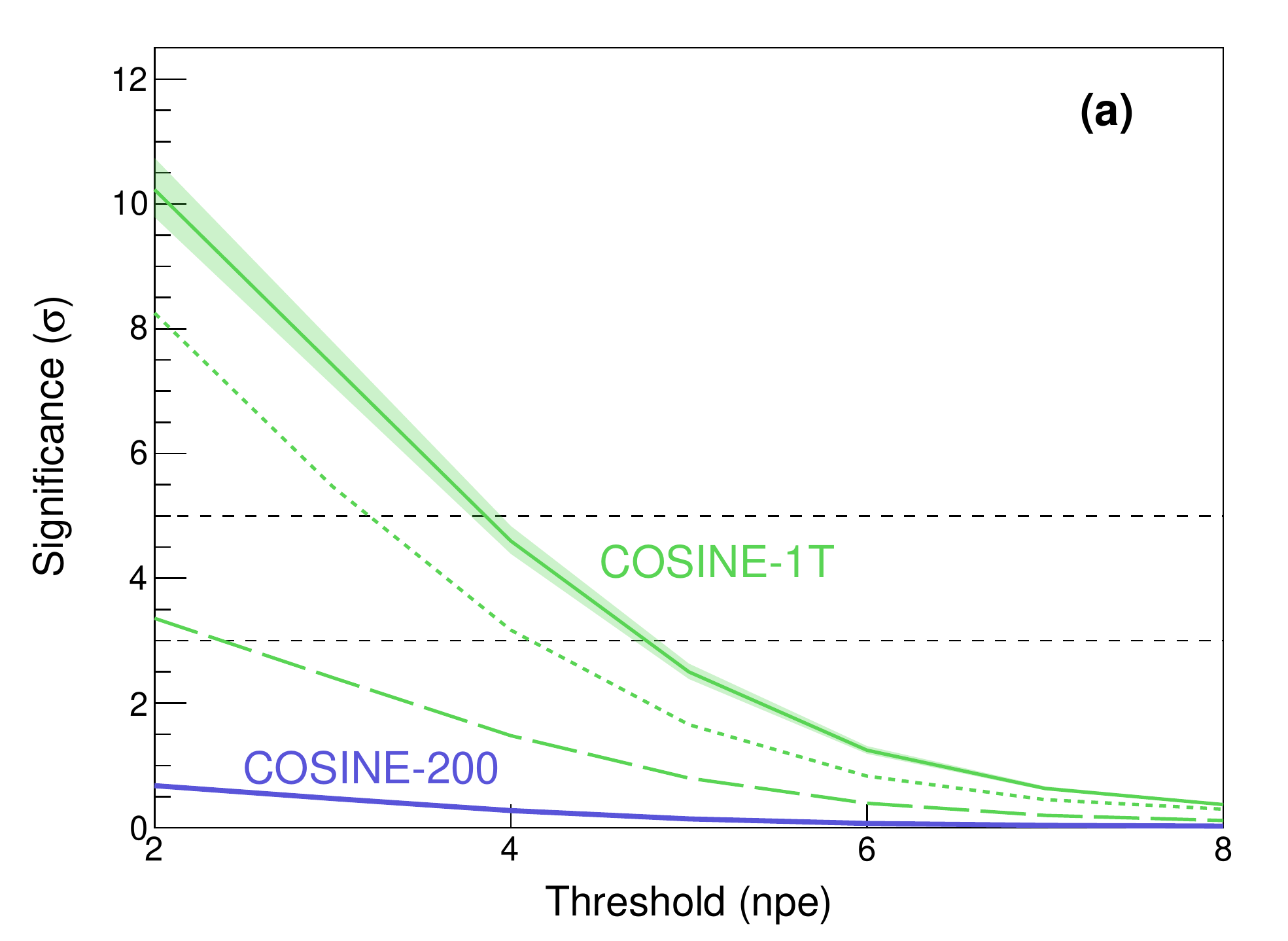} 
\includegraphics[width=0.49\columnwidth]{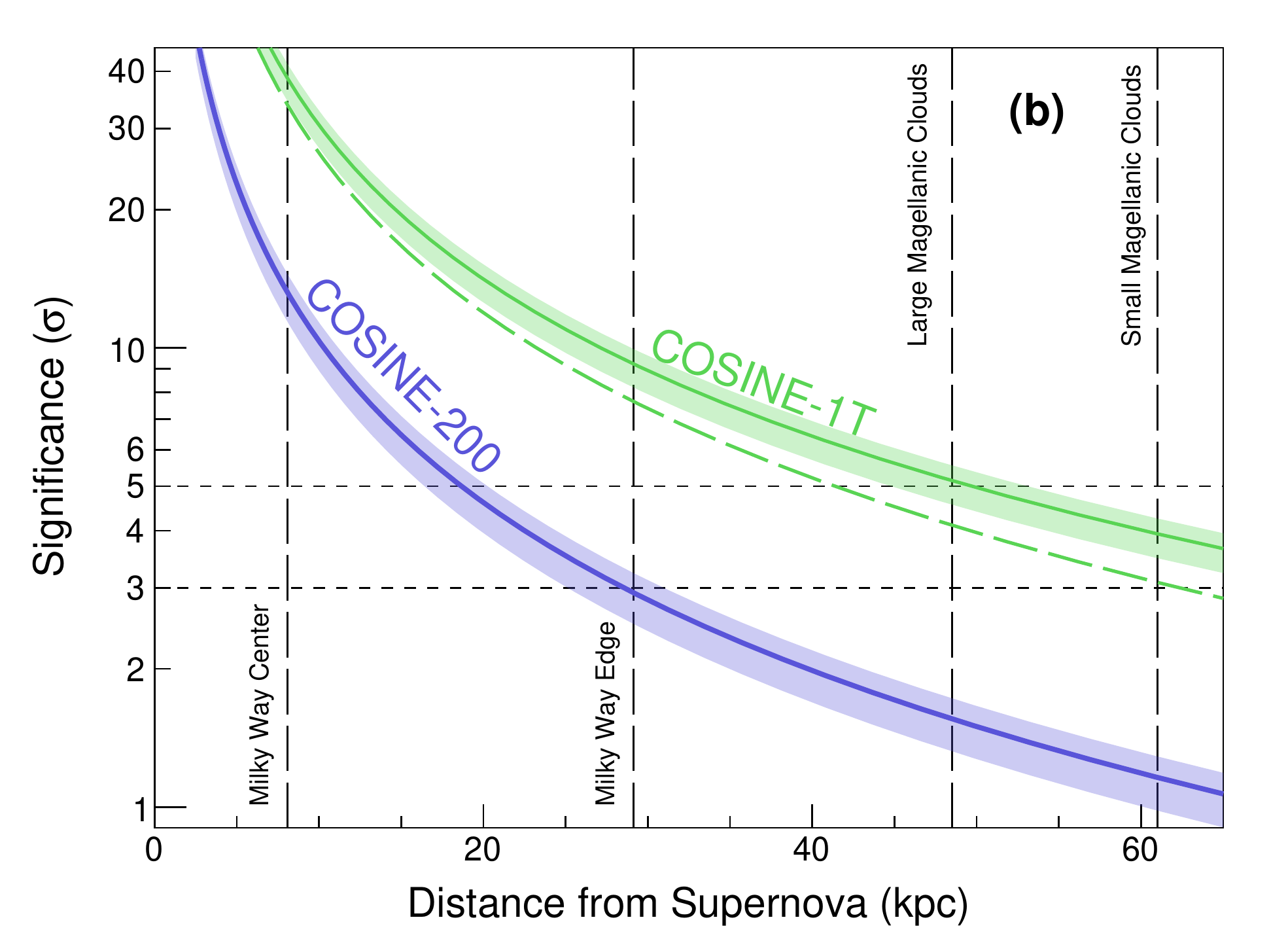} 
\caption{Sensitivity curves for 200-kg (blue) and 1000-kg (green) NaI(Tl) crystals. In case of 1000-kg NaI(Tl) sensitivities, we consider two different background scenarios of the same background of the COSINE-200 crystal (green-dashed line) and an order of magnitude reduced background from the COSINE-200 crystal (green-solid line). (a) Sensitivities for the solar-neutrino \cenns as a function of the analysis threshold are presented. Here the bands indicate the 10\% background systematic uncertainties. The green-dotted line shows the sensitivity for 1000-kg NaI(Tl) crystal assuming 50\% selection efficiency at the threshold region.  (b) Sensitivities for the supernova-neutrino \cenns regarding the distance to the supernova with an assumption of 4-NPE threshold are presented. The bands indicate the threshold variation from 2 to 6-NPE.}
\label{fig_sensi}
\end{center}
\end{figure*}

To estimate the sensitivity for various cases using the simplified method, as shown in Fig.~\ref{fig_sensi}, we employ the Asimov dataset~\cite{Cowan2011}, instead of multiple simulated experiments, using a simulated pseudo-dataset. The sensitivities are evaluated using Asimov data composed of the expected signals and backgrounds without statistical fluctuations as follows:
\begin{eqnarray}
n_{i,\mathrm{A}} = \mu's_i + b_i,
\end{eqnarray}
where $\mu' = 1$ to test for the presence of \cenns signals. Signal significance $S$ is calculated as $\sqrt{\chi^2} \sim\sqrt{t_0}$ in Eq.~\ref{eq:testS0},
\begin{eqnarray}
S = \sqrt{t_{0,\mathrm{A}}} = \sqrt{-2\ln \frac{L_\mathrm{A}(\mu=0)}{L_\mathrm{A}(\hat{\mu} = \mu')}},
\label{eq:significance}
\end{eqnarray}
where subscript A denotes  that Asimov data $n_{i,\mathrm{A}}$ are used.

Figure~\ref{fig_sensi} shows the significance for the observation sensitivities of \cenns signals from the (a) solar neutrino and (b) supernova neutrino, with systematic bands. As shown in Fig.~\ref{fig_sensi}~(a), the solar \cenns cannot be easily observed with COSINE-200, while one can observe the $>$3$\sigma$ significance through COSINE-1T with a 4--5-NPE threshold assuming 10 times reduced background from the COSINE-200 (green-solid line). If we cannot reduce the background from the COSINE-200 crystal, the soloar \cenns observation using the COSINE-1T detector is not easy (green-dashed line).  The effect of background systematic uncertainty is not large (green band), but the selection efficiency can reduce its significance (green-dotted line). 
Therefore, the background level reaching to the target of the COSINE-1T and high enough selection efficiency near energy threshold will be the key for the observation of the solar \cenns. 
By contrast, in case of supernova \cenns, as shown in Fig.~\ref{fig_sensi}~(b), the effect by background level is not significant due to the short exposure time (less than 10\,s) as shown in green-dashed line, and the observation sensitivities are high enough for the Milky Way supernovae. Even in the COSINE-200 experiment, supernovae within the Milky Way can be observed with $>$3$\sigma$ significance. 
Compared to liquid noble gas detectors~\cite{Baudis:2013qla,DARWIN:2020bnc,Lang:2015zhv,DarkSide20k:2020ymr}, the detector mass is much smaller and the expected background level is higher, so the expected sensitivities from the COSINE-1T are not compatible with  multi-ton liquid noble gas detectors. However, an observing \cenns neutrinos with different target material is important to understand \cenns process and to search for new physics~\cite{Abdullah:2022zue}. 

\section{Conclusion}
We investigated the prospects for measuring the \cenns from the solar and supernova neutrinos in future COSINE-200 and COSINE-1T NaI(Tl) dark matter search experiments. We obtained more than 3$\sigma$ observation sensitivities for supernova neutrinos from the COSINE-200 experiment. The COSINE-1T experiment may observe supernova neutrinos with above 3$\sigma$ significance as far as small magellanic clouds are concerned. However, the observation of solar neutrinos from future NaI(Tl) dark matter search experiments is marginal, reaching 3$\sigma$ sensitivities. Hence, significant improvement is required in the detector performance of the reduced background and increased light yields.

\section*{Acknowledgments}
This work was supported by the Institute for Basic Science (IBS) under the project code IBS-R016-A1, Republic of Korea.

\bibliography{dm}

\end{document}